# PECULIRIATIES OF TEMPERATURE DEPENDENCE FOR GENERALIZED HALL-PETCH LAW AND TWO-PHASE MODEL FOR DEFORMABLE POLYCRYSTALLINE MATERIALS

*A.A. RESHETNYAK[1]*
*Laboratory of Computer-Aided Design of Materials, Institute of\*
*Strength Physics and Materials Science of SB RAS, 634055 Tomsk, Russia*

## Abstract

In the framework of the suggested in [1] statistical theory of equilibrium flow stress, including yield strength, $\sigma_y$, of polycrystalline materials under quasi-static (in case of tensile strain) plastic deformation in dependence on average size, $d$, of the crystallites (grains) in the range, $10^{-8}$ m - $10^{-2}$ m it is found the coincidences of the theoretical and experimental data of $\sigma_y$ for the materials with BCC (α- Fe), FCC (Cu, Al, Ni) and HCP (α-Ti, Zr) crystal lattice at T=300K. The temperature dependence of the strength characteristics is studied. It is shown on the example of Al, that the yield strength grows with decreasing of the temperature for all grains with $d$ greater than $3d_0$ and then $\sigma_y$ decreases in the nano-crystalline region, thus determining a *temperature-dimension effect*. Stress-strain theoretical curves, $\sigma=\sigma(\varepsilon)$, are constructed for the pure crystalline phase of α- Fe with Backofen-Considére fracture criterion validity. The single-phase model of polycrystalline material is extended by means of inclusion of a softening grain boundary phase.

## Introduction

In the previous paper [1] the theory of equilibrium flow stress (FS) and, in particular yield strength, $\sigma_y$, was suggested on a basis of statistical approach to the quantized spectrum of the energy of plastic deformation of a grain (crystallite) of single-mode isotropic polycrystalline (PC) aggregate under quasi-static plastic deformation (PD), which are described by the relations:

$$\sigma(\varepsilon) = \sigma_0(\varepsilon) + \alpha m \frac{Gb}{d}\sqrt{\frac{6\sqrt{2}}{\pi} m_0 \varepsilon M(0)} \left(e^{M(\varepsilon)b/d} - 1\right)^{-\frac{1}{2}}, \quad M(\varepsilon) = \frac{Gb^3_\varepsilon}{2k_B T} = \frac{Gb^3}{2k_B T}(1+\varepsilon)^3, \quad (1)$$

with extreme grain size $d_0(\varepsilon,T)$ at PD value $\varepsilon$ and temperature $T$, for which the maximum of the FS $\sigma_m(\varepsilon)$ is reached,

$$d_0(\varepsilon,T) = b\frac{Gb^3(1+\varepsilon)^3}{2 \cdot 1.59363 k_B T}, \quad \sigma_m(\varepsilon) = \sigma_0(\varepsilon) + \alpha m G \sqrt{\frac{6\sqrt{2}}{\pi} m_0 \varepsilon \frac{b \cdot 1.59363}{d_0(\varepsilon,T)(1+\varepsilon)^3}} \left(e^{1.59363} - 1\right)^{-\frac{1}{2}}. \quad (2)$$

In the relationships (1), (2) the polyhedral parameter $m_0$ [1] is related with the Hall-Petch (HP) coefficient $k(\varepsilon)$ in the normal HP law for $\varepsilon = 0,002$, which follows from (1) in the coarse-grained (CG) limit by the formula

$$\sigma(\varepsilon)|_{d \gg b} = \sigma_0(\varepsilon) + k(\varepsilon)d^{-\frac{1}{2}}, \Rightarrow m_0 = \frac{\pi}{6\sqrt{2}} \frac{k^2(\varepsilon)}{(\alpha m G)^2 \varepsilon b} \frac{M(\varepsilon)}{M(0)}, m_0 \cdot \alpha^2 = \left(\frac{k(\varepsilon)}{mG}\right)^2 \frac{\pi}{6\sqrt{2}\varepsilon b}(1+\varepsilon)^3 \quad (3)$$

that one allows to refine the values of $m_0$ in accordance with experimental data for $k(\varepsilon)$ in various CG materials and to formulate the model in terms of the constant $(m_0 \cdot \alpha^2)$, now in whole region of diameter $d$.

We introduce in the paper a quasi-particle interpretation for the quanta of PD energy to be equal to the energy $E_d^{L_e}$ of unit dislocation to be necessary for appearance of a dislocation (or nanopore within suggested in [1] scenario of emergence of a edge dislocation from sequence of 0D-defects, defining a zone of localized plasticity, or band of localized deformation). Second, we check with use of plots the validity of the generalized HP law for the yield strength of a number of single-mode PC materials with different crystal lattice (CL). Third, we study a temperature dependence of $\sigma_y$, thus, revealing the *temperature-dimension effect*, on the example of $Al^2$. The obtained relationship (1) permits to construct stress-strain curves, $\sigma = \sigma(\varepsilon)$, for different sizes of grain of single-mode homogeneous PC aggregate with only the single (crystallite) phase and study its hardening and pre-fracture stages. Finally, the inclu-sion of the second (grain boundary) phase leads to the construction of a realistic model permitting to control the defect structure of both phases. The paper is devoted to the solutions of the problems above.

---

[1] e-mail: reshet@ispms.tsc.ru

[2] Note, e.g. the dependence of $k(\varepsilon) = f(T)$ were considered for *Pb* in [2] and for another materials see the review [3])

We use the notations and conventions of the paper [1].

**Quasi-particle interpretation of the crystallite energy quantization under plastic deformation**

In the framework of Louis de Broglie corpuscular-wave hypothesis the minimal part of energy, $E_d^{L_e}$, of PD should possesses by the properties of a particle and wave. The latter means that the energy of such quasi-particle, conditionally called as "*dislocon*", (as the quantum of PD energy): $\hbar\omega = E_d^{L_e} + W$, for creating (or changing) of 0*D* (zero-dimensional) or 1*D* defect cannot be less than $\hbar\omega_{red} = E_d^{L_e}$, thus determining the "red" border of frequency. For instance, for α-Fe it equals for ε=0: $\omega_{red}(\alpha\text{-Fe})=5,99\cdot 10^{15} s^{-1}$ and for a lower frequency a *dislocon* in α-Fe does not arise, therefore not generating (or not changing) a generalized dislocation [1] in the crystallite. Because of, a *dislocon* plays the role of a carrier of interaction among dislocations in the crystallite, let us choose a dispersion law: frequency dependence (momentum *p*) from a wave vector *k*: $\omega = \omega(k)$ in the linear form as for massless particles being subject to the Bose-Einstein statistic (as for the acoustic phonons in Debye approximation):

$$\omega = v_d k \ (p = \hbar k) \ \text{and for} \ k_{max} = \pi/b_\varepsilon \Rightarrow v_d = \omega/k_{max} = \omega b_\varepsilon/\pi \ , \tag{4}$$

with the speed of propagation of dislocon in the medium: $v_d$, being evaluated for α-Fe at ε=0 as $v_d$(α-Fe)=$5,18\cdot 10^5 m/s$ (on the first Brillouin zone border). The value of $v_d$ is more than on 2 orders of magnitude as the speed of sound ($v_s$=5,93 $\cdot 10^3$ м/с), to be related to harmonic (phonon) oscillations of CL, but not to its local destruction. However, as we will show, the choice: $v_d = M(0)v_s$ makes by closed to the correct one the interpretation of the *dislocon* as the composite quasi-particle consisting from acoustic phonons, which were formed when the bond between atoms is broken (e.g., among A and A´ on the Fig. 2a in [1] when generating of a nanopore).

At the thermodynamic quasi-equilibrium state for fixed value $\varepsilon^3$ of residual PD in each crystallite the quasi-equilibrium process of emission and absorption of the dislocons is established (see Ref. [1] for explanation), meaning that in the crystallite the standing waves from acoustic phonons are placed. The relation (4) means, that in the crystallite the plane wave in the crystallographic plane is propagated along axis *z*:

$$u(r) = u_0 \exp[i(\omega t + kz)], \tag{5}$$

with periodic boundary conditions: $\exp[ikz] = \exp[ik(z+d)]$ being valid for $k\frac{2\pi n}{d}, n \in Z$. Hence, on the length of deformable part of the volume: $L_\varepsilon = 2\pi/(\varepsilon d(1+\varepsilon))$, in *k*-space there is one allowed value of *k* and the number of modes in the unit of *k*-length is equal to: $L_\varepsilon/2\pi$. The total number of modes bounded in the limits of thin ring from the radius *k* to radius (*k*+*dk*), with account for the pairing of the dislocations and polyhedral parameter $m_0$ is evaluated as:

$$dn = 2m_0 \cdot 2\pi[k + dk - k]\frac{\varepsilon d(1+\varepsilon)}{2\pi} = 2m_0 \frac{\varepsilon d(1+\varepsilon)}{v_{3e}}d\omega = 2m_0 \frac{M(0)\varepsilon d(1+\varepsilon)}{v_d}d\omega, \tag{6}$$

where it was taken into account that dislocons (phonons) are localized only in the zone of localized deformation, i.e. in the plastically deformed part of linear volume: ($L = \varepsilon d(1+\varepsilon)$). The only those from them, for which the frequency: $\omega \geq \omega_{red}$ give the possibility of creation of a dislocation (nanopore). The bandwidth of the frequencies for such dislocons should be extremely narrow: $\omega(\varepsilon) - \omega_{red} \ll \omega_{red}$ and in the framework of Einstein proposal we naturally assume for the dislocons, that all frequencies for them coincide with $\omega_{red}$, that corresponds to the insertion of Dirac $\delta$-function in (6): $(\pi^{-1}\omega \cdot \delta(\omega - \omega_{red}))$:

$$N^* = \int_0^\infty \frac{2\, m_0 \varepsilon d(1+\varepsilon)}{\pi v_s} \omega \cdot \delta(\omega - \omega_{red}) d\omega = \frac{2 m_0 M(0) \omega_{red}\, \varepsilon d(1+\varepsilon)}{\pi v_d} = \frac{2 m_0 M(0) \varepsilon d(1+\varepsilon)}{b(1+\varepsilon)}. \quad (7)$$

---

[3] In fact, at constant quasi-static deformation rate there is no thermodynamic equilibrium state for any of the crystallites in the PC sample, but we have argued in the footnote 4 in [1] in the item 4 of definition of a probability space ($\Omega, U, P$) in [1] that the process of PD may be presented in the form of the sequence of equilibrium processes being changed at changing of ε by hopping from one, ε, to another, ε+Δε, so that the probability distribution of $P(E_n, \varepsilon)$, for any of the possible defects in an elementary PD act to occur in a crystallite, depends smoothly on the strain ε: $P(E_n, \varepsilon) = f(\varepsilon)P(E_n, 0)$ such that $P(E_n, \varepsilon_1) > P(E_n, \varepsilon_2)$ for $\varepsilon_1 > \varepsilon_2$.

An allowance for the model isotropy of crystallite distribution in the PC sample leads to the coincidence of (7) with number of dislocations $N^*/\sqrt{2} = N_0/\sqrt{2}$ for value of PD $\varepsilon$, earlier obtained from mechanical arguments in [1].

## Implementation of the generalized Hall-Petch law for α-Fe, Cu, Al, Ni, α-Ti, Zr.

In order to determine the values of the constant $m_0$ (3) let us use the known experimental values for HP coefficient $k(0,002)$ for (poly)crystalline single-mode samples with BCC, FCC and HCP CL from the Table 1 with corresponding values for $\sigma_0$, $G$, lattice constants $a$ [4], Burgers vectors with the least possible lengths $b$, with respective most realizable sliding systems (given in the Table 2), constant of interaction for the dislocation $\alpha$ [5, 6] and computed values of the least unit dislocations $E_d^{L_e}$, extreme grain sizes $d_0$, maximal differences of $\sigma_y$ in accordance with (5) in [1] and (2) for $T=300$K:

| Type of CL | BCC | FCC | | | HCP | |
|---|---|---|---|---|---|---|
| Material | α-Fe | Cu | Al | Ni | α-Ti | Zr |
| $\sigma_0$, MPa | 170 (annealed) | 70 (anneal.); 380 (cold-worked) | 22 (anneal. 99,95%); 30 (99,5%) | 80 (annealed) | 100(~100%); 300 (99,6%) | 80-115 |
| $b$, nm | $\frac{\sqrt{3}}{2}a$ =0,248 | $a/\sqrt{2}$ =0,256 | $a/\sqrt{2}$ =0,286 | $a/\sqrt{2}$ =0.249 | $a$=0,295 | $a$=0,323 |
| $G$, GPa | 82,5 | 44 | 26,5 | 76 | 41,4 | 34 |
| $T$, K | 300 | 300 | 300 | 300 | 300 | 300 |
| $k$, MPa·$m^{1/2}$ | 0,55-0,65 ($10^{-5}-10^{-3}m$) | 0,25 ($10^{-4}-10^{-3}m$) | 0,15 ($10^{-4}-10^{-3}m$) | 0,28 ($10^{-5}-10^{-3}m$) | 0,38-0,43 ($10^{-5}-10^{-3}m$) | 0.26 ($10^{-5}-10^{-3}m$) |
| $\alpha$ | – | 0,38 | – | 0,35 | 0,97 | – |
| $E_d^{L_e} = \frac{1}{2}Gb^3$, eV | 3,93 | 1,28 | 1,96 | 3,72 | 3,33 | 3,57 |
| $m_0 \cdot \alpha^2$ | 3,66-5,11 | 2,57 | 2,28 | 1,11 | 5,83-7,47 | 3,69 |
| $d_0$, nm | 23,6 | 14,4 | 13,6 | 22,6 | 23,8 | 28,0 |
| $\Delta\sigma_m$, GPa | 2,14-2,56 | 1,27 | 0,81 | 1,12 | 1,49-1,70 | 0,90 |

Table 1: Values $\sigma_0$, $\Delta\sigma_m = (\sigma_m - \sigma_0)$, $E_d^{L_e}$, $k$, $m_0$, $\alpha$ for BCC, FCC and HCP polycrystalline metal samples with $d_0$, $b$, $G$ obtained with use of Ref. [4] at $\varepsilon = 0,002$.

The values for $k$ at $\varepsilon = 0,002$ are taken for α-Fe, Cu, Ni from [3], Zr from [6] for Al from [4], for α-Ti from [7, 8] in the range for the grains shown in the brackets.

| | α-Fe | Cu | Al | Ni | α-Ti | Zr |
|---|---|---|---|---|---|---|
| Plain | {110}, {112}, {123} | {111} | {100},{111} | {111} | $(10\bar{1}0), (10\bar{1}1), (0001)$ | $(10\bar{1}0)$ |
| Direction | <111> | <110> | <110> | <110> | $[2\bar{1}\bar{1}0]$ | $[11\bar{2}0]$ |

Table 2: The most probable sliding systems at $T=300$K [6] for α-Fe, Cu, Al, Ni, α-Ti, Zr in terms of the Miller indices for BCC, FCC and Miller-Bravais indices for HCP lattices.

The graphical dependences $\sigma_y = \sigma_y(d^{-1/2})$ for crystallite phase of the polycrystalline aggregates of α-Fe, Cu, Al, Ni, α-Ti, Zr with closely-packed randomly oriented grains being homogeneous with respect to its size (single-modesamples) at T=300K are presented on the Fig. 1 on a basis of the Tables 1, 2:

According to Fig. 5, experimental data coincide approximately at the extreme size values [3], as well as the values for maximums $\sigma_m$ [16]. The values of $d_0(0.002, 300)$, e.g., for α-Fe, Cu, Al, Ni, α-Ti, Zr given in Table 1, is in complete agreement with the range (both empirical and theoretical) of critical size values for the average diameters of grains $d_{cr}$ for PC samples (listed, e.g., in Ref. [3] (Table 2.6, pp. 110–111), ranging from 5–10 nm to 20–50 nm, particularly, for $d_{cr}$(Cu)=10 nm≈ $d_0$(Cu)=14.4 nm from Ref. [17] and $d_{cr}$(Ni)=20 nm ≈ $d_0$(Ni)=22.6 nm. For the corresponding values for the maxima of experimental $\sigma_y$, i.e., $\bar{\sigma}_m(0.002)$ and $\sigma_m(0.002)$ in single-mode (in average) PC samples (see, e.g., Refs. [16], [18]), we find that $\bar{\sigma}_m(0.002)$ (α-Fe) ≈ 2.75 GPa, $\bar{\sigma}_m(0.002)$ (Ni) ≈ 1.7 GPa, $\bar{\sigma}_m(0.002)$ (Cu) ≈1.0 GPa coincide approximately with the theoretical maxima, with allowance for the various definitions of HP coefficients according to the literature (see Table 2 in [3] with $k(Cu) \in [0.01, 0.024]$ for UFG PC samples). The difference for Ni and Cu may be explained by leaving out of account, first, softening due to weak grain boundary parts, which leads to a decrease in $d_0$, $\sigma_m$, and, second, excitation at PDs of other dislocation ensembles, especially in the NC region with a Burgers vector $b_1$ larger than the one for the most probable dislocation, and therefore with a larger unit dislocation energy, due to $E_d^{Le}(G,b) < E_d^{Le}(G,b_1)$, and a larger input to $\sigma(0.002)$, as for Ni. Therefore, the maximum values $\sigma_m$ demand taking into account a negative input from the GB phase.

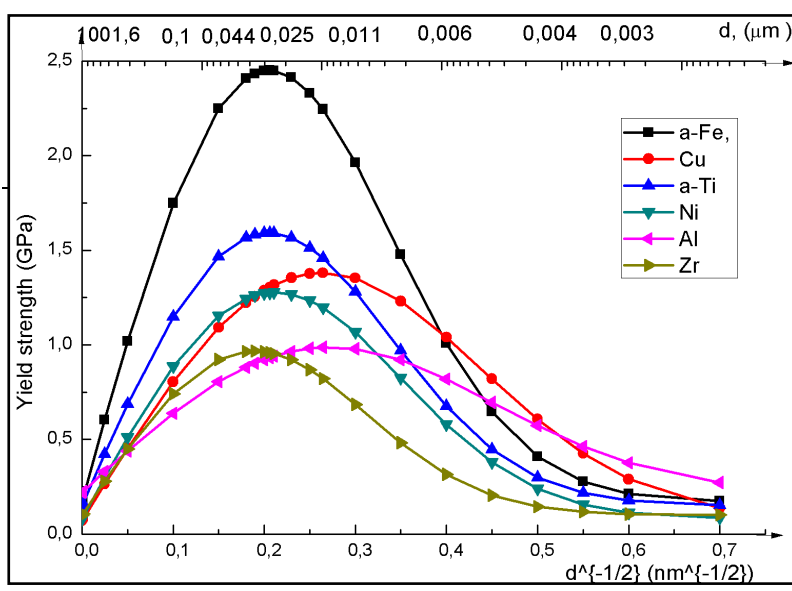

Fig. 1. Graphical dependences (plots) for generalized Hall-Petch law (1) at $\varepsilon = 0{,}002$ with additional upper scale with size of the grains $d$ given in μm. Upper axis $d$ is changing within range $(\infty;0)$ with inverse quadratic scale and correspondence: (100; 1,6; 0,1; 0,044; 0.025; 0,011; 0,006; 0,004; 0,003) μm ↔ (0,005; 0,015; 0,1; 0,15; 0,2; 0,3; 0,41; 0,5; 0,57) $nm^{-1/2}$ for respective values on lower axis.
The least from possible values of the parameters $m_0(k)$ for α-Fe, α-Ti, values of $\sigma_0$ for annealed materials with maxima for $\sigma_y$ are calculated in the respective to the Table 1 extreme grain sizes $d_0$.

### Temperature dependence of yield strength and extreme grain sizes for Al.

Because of with a growth of the temperature the value of the shear module $G(T)$ (as well as $\sigma_0(T)$) decreases, whereas the linear parameters $b$ and $d$ are increased with the same true linear coefficient of the temperature expansion $\alpha_d$ [4] (for BCC and FCC materials), then the extreme grain size $d_0(\varepsilon)$ is shifted in to region of smaller grains: $d_0(\varepsilon,T) > d_0(\varepsilon,T')$ for $T' > T$; $T',T \in [T_1,T_2]$ (within the same phase of the material [1]) according to the rule:

$$d_0(\varepsilon,T') = b_\varepsilon(T') \frac{\frac{1}{2}G(T')[b_\varepsilon(T')]^3}{1{,}59363 \cdot kT} = d_0(\varepsilon,T)g(\alpha_G,\alpha_d,T,T'),$$
$$g(\alpha_G,\alpha_d,T,T') = \left(\frac{b_\varepsilon(T')}{b_\varepsilon(T)}\right)^4 \frac{G(T')T}{G(T)T'} = (1+\alpha_d(T'-T))^4(1-\alpha_G(T'-T))\frac{T}{T'}, \quad (8)$$
$$\text{for } b_\varepsilon(T') = b_\varepsilon(T)(1+\alpha_d(T'-T)); \quad G(T') = (1-\alpha_G(T'-T))G(T)$$

with linear temperature coefficient for the shear modulus $\alpha_G$, e.g. for Al: $\alpha_G = 5{,}2 \cdot 10^{-4} K^{-1}$ being approximately constant within temperature range [250K,300K]. It follows from (8) that for varying of $T$ in a small range the value $d_0(\varepsilon,T)$ is multiplicatively changed with the factor $g(\alpha_G,\alpha_d,T,T')$. At the same time with accommodation of PD $\varepsilon$ the value of $d_0(\varepsilon,T)$ is shifted to the area of larger grains: $d_0(\varepsilon_1,T) > d_0(\varepsilon_2,T)$ for $\varepsilon_1 > \varepsilon_2$. Both behavior of $\sigma(\varepsilon)$ and $\sigma_m(\varepsilon)$ for a monotonic change of the temperature is added from $T$- behavior of the crystal and dislocation substructures of the crystallite phase of the sample. The quantities $(\sigma_0, G) = (\sigma_0(T), G(T))$ correspond to the former substructure being decreased and $(b, d)$ being increased with grows of $T$, whereas for the latter one it corresponds the explicit dependence from temperature in the relations (1), (2), that leads to increasing of $\sigma, \sigma_m$ with grows of $T$. A set of systematical experimental data concerning the temperature dependence research of $\sigma(\varepsilon)$, $\sigma_0(\varepsilon), d_0(\varepsilon)$ *is absent in the literature* and it is necessary to replenish it. For instance, the results of molecular dynamics simulation for Cu [9] give contradictory data that with the grows of $T$ the values of $\sigma_y$ and its maximum $\sigma_m(\varepsilon_{0,2})$ is decreased for any fixed $d$, as well as the simulated values of $d_0(\varepsilon_{0,2})$ are shifted to the area of larger grains, from 4nm at T=280K to 25nm at T=370K. The latter results, (obtained under high-speed mechanical loading with $\dot\varepsilon. = 5 \cdot 10^8\, s^{-1}$) contradicts to the established displacement law (2). However, one should be stressed, that the derivation of (2) is based on the assumption of exist-ence of the equilibrium probabilities distribution: $P_n(\varepsilon) = A(\varepsilon)e^{-\frac{\frac{1}{2}Gb_\varepsilon^3}{k_B T}\frac{E_n}{E_N}}$ given by the Eq.(10) in [1], introduced for the case of quasi-static loading in the state of thermodynamic quasi-equilibrium for given $\varepsilon$, which plays the role of adiabaticity parameter. For the maximal differences: $\Delta_m \sigma(\varepsilon, T) = \sigma_m(\varepsilon, T) - \sigma_0(\varepsilon, T)$ calculated for $T' > T$ its ratio is given by:

$$\frac{\Delta_m \sigma(\varepsilon, T)}{\Delta_m \sigma(\varepsilon, T')} = \frac{G(T)}{G(T')}\sqrt{\frac{b_\varepsilon(T)d_0(\varepsilon, T')}{b_\varepsilon(T')d_0(\varepsilon, T)}} = (1 + \alpha_G(T'-T))\sqrt{(1 - \alpha_d(T'-T))}g(\alpha_G, \alpha_d, T, T') < 1, \quad (9)$$

and in view of the approximation for the right-hand side $r(\varepsilon, T, T') = \{1 + (0,5\alpha_G + 1,5\alpha_d)(T'-T)\}\sqrt{T/T'}$, where a decreasing of the root, $\sqrt{T/T'}$, suppresses the grows of the first multiplier, e.g. for Al in the limits $[T, T'] = [300, 350]K$ with $\alpha_d(Al) = 2,33 \cdot 10^{-5} K^{-1}$ [4] the estimation $\Delta_m \sigma(\varepsilon, T)/\Delta_m \sigma(\varepsilon, T') = 0,93$ holds.

An experimental estimation of the grows $\sigma(\varepsilon, T) > \sigma(\varepsilon, T')$ in this case means that the stress $\sigma_0(\varepsilon, T)$ should decrease more than increase of $\Delta_m \sigma(\varepsilon, T)$ with growth of the temperature. However, the value of $\sigma_0(\varepsilon, T)$ may range from 7% to 20 % from $\sigma_m(\varepsilon, T)$ when $T=300K$ for various materials (see Table 3 below). In the low-temperature region the value of $\sigma_y(d,T)$ significantly increases, that it is explained in [6] by a dominance of twinning in BCC, FCC and especially in HCP metal poly-crystals. Note, the latter process (twinning) does not contribute into plasticity, but provides an appearance of additional sliding systems for dislocations (due to the changing of curvature-torsion of CL). For the process of twinning in the CG-region of PC materials have been obtained an analogous to the normal form of HP law (3) relationship for $\sigma_y(d,T)$ but with own $\sigma_{0tw}(T)$ and HP coefficient $K_{tw}(T)$, such that, the first quantity being less than for one for pure dislocation $\sigma_0(T)$, and with the second quantity being multiply large (in 5 times [6] for Cr), than for $k(\varepsilon)$. The question of description of an input from the twinning into the deformation hardening together with further research of $T$-dependence for $\sigma_y(d,T)$ we leave out of the paper's scope. However, we note that twinning (as forming of 2D-defects) may be as well considered in terms of the clusters of partial dislocations (see chapter 23 in [10] for details)![4] The condition of decreasing almost everywhere of $\sigma(\varepsilon, T)$ with a grows of $T \in [T_1, T_2]$, is equivalent to the condition:

$$\frac{d\sigma}{dT} = \frac{d\sigma_0}{dT} + \alpha m \frac{Gb}{2d}\left\{\left(e^{M(\varepsilon)b/d} - 1\right)\frac{d}{dT}\ln\left(\frac{G^2}{M(0)}\right) - \frac{b}{d}e^{M(\varepsilon)b/d}\frac{dM}{dT}\right\}\sqrt{\frac{6\sqrt{2}}{\pi}m_0\varepsilon(1+\varepsilon)M(0)}\left(e^{M(\varepsilon)b/d} - 1\right)^{-\frac{3}{2}} \leq 0. (10)$$

For CG and fine-grain (FG) materials $(b \ll d)$ equation (1) takes form of normal HP law (3), being valid as well for $\varepsilon > 0,002$ in accordance with [5]:

$$\sigma(\varepsilon)|_{d \gg b} = \sigma_0(\varepsilon) + k(\varepsilon)d^{-\frac{1}{2}}, \quad k(\varepsilon) = \alpha m G \sqrt{\frac{6\sqrt{2}}{\pi}m_0\varepsilon b \frac{M(0)}{M(\varepsilon)}}, \quad \frac{M(0)}{M(\varepsilon)} = (1+\varepsilon)^{-3}, \quad (11)$$

from which in view of $T$-independence of the relation $b/d$ it follows, that with grows of $T$, the yield strength $\sigma_y(T)|_{d \gg b} = \sigma(0,002; T)|_{d \gg b}$ decreases due to decreasing of $\sigma_0$ and $G$. Outside from GC, FG and ultra-FG (UFG) regions the FS has opposite temperature behavior.

Thus, on the 3-dimensional graph $(d, T, \sigma(\varepsilon, d, T))$ of the equation (1) there exist the value $d_1(\varepsilon)$[5] $(d_1 \approx 3d_0 \gg b)$, such that for all $d > d_1(\varepsilon)$ the stress $\sigma(\varepsilon, d, T)$ decreases with grows of T, whereas for $d < d_1(\varepsilon)$ the value of $\sigma(\varepsilon,d,T)$ increases (meaning the presence of the maxima for $\sigma(\varepsilon,d,T)$ with respect to average size of the crystallites and maxima with respect to $T$ at fixed $\varepsilon$). For low $T$ the model should provide a significant grows of $\sigma(\varepsilon,d,T)$ and requires a clarification, however for $d \gg bM(\varepsilon)$ a value of FS is determined in view of (11) by the behavior of $G$ and by little-studied behavior of $\sigma_0(T)$.

---

[4] Therefore, we may think that analytically the twinning influence into the deformation hardening is already taken into account the dislocations forming them.

[5] An approximate estimation for $d_1(\varepsilon)$ follows from the cubic generalization of HP law (3), (11), being obtained from the representation (1), by the retention of the quadratic terms in powers of $b/d$ from $\left(e^{M(\varepsilon)b/d} - 1\right)$ in the form: $\sigma(\varepsilon)|_{d \gg b} = \sigma_0(\varepsilon) + k(\varepsilon)d^{-1/2}(1 - \frac{1}{4}M(\varepsilon)b/d)$. From the extremity of $\sigma(\varepsilon)|_{d \gg b}$ with respect to $T$ for omitting $\frac{d\sigma_0}{dT}$ and leaving of the leading term with an explicit dependence from $T$ in $M(\varepsilon)$ it follows, that $d_1 = \frac{bM(\varepsilon)}{4\alpha_G}\left(\frac{1}{T} + \alpha_G\right)$. For Al at $T=300K$ we have $d_1 = 138b = 39,6$ nm. Thus, $d_1 \approx 3d_0(\varepsilon, T)$.

One expect an analogous $T$-dependence for $\sigma_0(T)$, as one for $G$ as the quantities characterizing a mono-crystal. Note, for HP coefficient, that, first, $k(\varepsilon)$ in (3) and in (11) determined in the CG limit does not explicitly depend from the temperature (see, for chromium Fig..2.16 in [6] up to 100K), second, there exists up to 10 analytical definitions for $k(0,002)$ [3] (see as well the Table 5 in [6]).

The temperature dependence of the yield strength $\sigma_y(d,T)$ in the coordinates $(d^{-1/2}, \sigma_y)$, for instance, for Al presented in the parametric form on the Fig. 2 with assumption of the same dependence for the parameter $\sigma_0(T)$ as for $G(T)$ (for instance, with coefficient $[1 - 5,2 \cdot 10^{-4}(T'-T)]$ as in (8) in view of absence of the experimental data).

| T,K \ Al | G, GPa | $\sigma_0$, MPa | $d_0$, nm | $(\sigma_m-\sigma_0)$, GPa |
|---|---|---|---|---|
| 350 | 25,8 | 21 | 11,3 | 0,85 |
| 300 | 26,5 | 22 | 13,6 | 0,81 |
| 250 | 27,4 | 23 | 16,8 | 0,74 |
| 200 | 28,1 | 23,5 | 21,5 | 0,67 |
| 150 | 28,8 | 24 | 29,5 | 0,59 |

Table 3 The values of parameters $\sigma_0, d_0, G, (\sigma_m-\sigma_0)$ for Al in the temperature range [150,350]K.

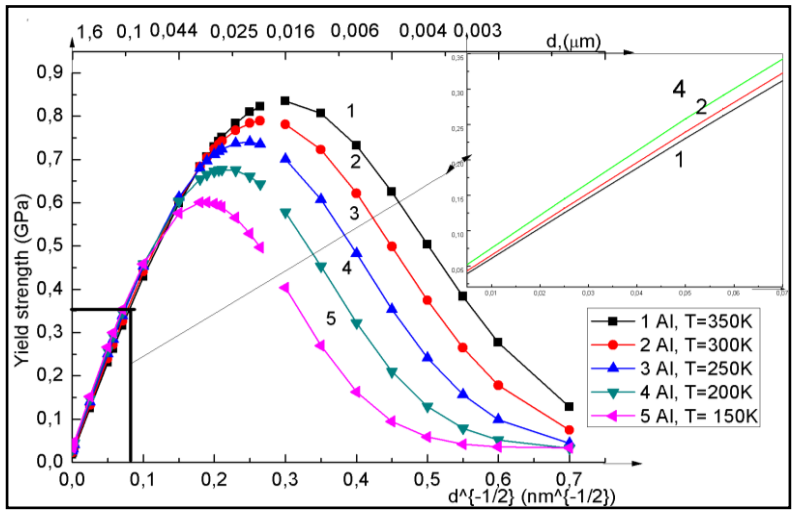

Fig. 2. Graphical dependences (plots) for generalized Hall-Petch law (1) at $\varepsilon = 0,002$ for Al at $T$=350; 300; 250; 200; 150$K$. On the input the part of the dependences in CG and FG regions is shown (extracted as the rectangle) for $T$=350K; 300K; 200K with normal HP law (11) validity. The straights (3) for $T$=250K and (5) for $T$=150K are situated on the input respectively between the straights (2), (4) and above straight (4).

From the Fig. 2 it follows that in the range of grains up to 0,1 μm the $T$-behavior of $\sigma_y(d,T)$ has usual form ($\sigma_y$ decreases with grows of $T$), whereas both for $d<d_1$=39,6 nm when going to anomalous part of HP law the $T$-behavior of $\sigma_y(d,T)$ looks as unusual ($\sigma_y$ increases with grows of $T$). In particular, for $d=d_1$ the yield strength $\sigma_y(d_1,T)$ does not depend upon $T$ in the sufficiently wide range of temperatures. Indeed, for instance, for $T$=150 K extreme for the density of dislocations and yield strength are reached at $d_0(150)$=29,5 nm and then with decreasing of $d$, say down to $d$=21 nm, dislocations leave the bodies of grains and the softening occurs so that $\sigma_y(29,5;150) > \sigma_y(21;150)$. When such a stressed sample of $Al$ is heated up to $T$=250 K this grain with $d$=29,5 nm is even pre-extreme (29,5>$d_0(250)$=16,8 nm), and therefore with decreasing of $d$ down to $d$=21 nm, under plastic deformation, the scalar density of dislocations increased even more together with the increase of the yield strength and an inverse inequality is valid: $\sigma_y(29,5;150) < \sigma_y(21;150)$. The such behavior is also expected for arbitrary PC materials with tightly packed grains and should be experimentally checked (with allowance for a possible change of grains be-cause of recrystallization process, especially for *Al*). We refer to this phenomenon as a *temperature-dimension effect* (TDE) in PC materials, which is characterized by the two following properties, at least in a sufficiently wide range of $\varepsilon$ under plastic deformations:

1) a displacement of the extremal size value $d_0(\varepsilon, T)$ (22) to the large grains region[6] with a decreasing of the temperature, $d_0(\varepsilon, T_1) > d_0(\varepsilon, T_2)$ for $T_1 < T_2$;

2) an increase of FS $\sigma(\varepsilon)$, including the maximum $\sigma_m(\varepsilon)$ (23), with a grows of T in the NC region in single-mode PC materials for $d < d_1 \approx 3d_0$, and a decrease for $d > d_1$.

An including of the second GB phase in the model may significantly change this effect.

## Stress-strain curves for crystallite phase of α-Fe. Backofen-Considére criterion.

A dependence of $\sigma(\varepsilon)$ (1) together with stress-strain curve plot $\sigma = \sigma(\varepsilon)$ (Fig. 3) permits one to find strain hardening coefficient $\theta(\varepsilon) = d\sigma/d\varepsilon$ (with assumption that $\theta_0(\varepsilon) = d\sigma_0/d\varepsilon$):

$$\theta(\varepsilon) = \theta_0(\varepsilon) + \alpha m \frac{Gb}{d\sqrt{\varepsilon}} \sqrt{\frac{3}{\pi\sqrt{2}}} m_0 M(0) \left\{ e^{M(\varepsilon)b/d} - 1 - \frac{3\varepsilon}{1+\varepsilon} M(\varepsilon) \frac{b}{d} e^{M(\varepsilon)b/d} \right\} \left( e^{M(\varepsilon)b/d} - 1 \right)^{-\frac{3}{2}}. \quad (12)$$

The stress-strain curves of the dependence $\sigma = \sigma(\varepsilon, d, T)$ (1) for the pure crystalline phase of the equilibrium PC α-Fe sample at *T*=300K for various average grain sizes are shown on the Fig. 3, on a basis of the Table 4. The values of conditional elastic limit $\sigma(0,0005)$ are formally calculated according to (1).

| α-Fe | $\sigma(\varepsilon)$, GPa. | | | | | | | | | |
|---|---|---|---|---|---|---|---|---|---|---|
| $\varepsilon \times 10^{-2}$ / $d, nm$ | 0,05 | 0,1 | 0,2 | 0,5 | 1 | 2 | 5 | 10 | 20 | 30 |
| $10^6$ | 0,009 | 0,012 | 0,017 | 0,027 | 0,038 | 0,054 | 0,081 | 0,107 | 0,133 | 0,144 |
| $10^5$ | 0,028 | 0,039 | 0,055 | 0,086 | 0,122 | 0,169 | 0,256 | 0,338 | 0,420 | 0,456 |
| $10^3$ | 0,273 | 0,386 | 0,545 | 0,856 | 1,204 | 1,677 | 2,536 | 3,339 | 4,129 | 4,465, |
| 150 | 0,667 | 0,943 | 1,332 | 2,095 | 2,938 | 4,086 | 6,147 | 8,016 | 9,690 | 10,197 |
| 23,6 | 1,145 | 1,617 | 2,279 | 3,572 | 4,976 | 6,827 | 9,840 | 11,859 | 11,861 | 9,834 |
| 10 | 0,827 | 1,166 | 1,640, | 2,548 | 3,500 | 4,666 | 6,140 | 6,254 | 4,192 | 2,130 |

Table 4 The values of the stress $\sigma(\varepsilon)$ for the crystalline phase of a single-mode polycrystalline α-Fe at *T*=300K for various average grain sizes in the range $\varepsilon \in [0,0005;0,3]$.

The reaching of the maximal value $\sigma_{\max}(\varepsilon)$ follows from the condition: $d\sigma/d\varepsilon = 0$, analogous to one given by (2):

$$\left( e^{y(\varepsilon)} - 1 \right)^{-\frac{3}{2}} / 2\sqrt{\varepsilon} \left\{ e^{y(\varepsilon)} \left( 1 - \frac{3\varepsilon}{1+\varepsilon} y(\varepsilon) \right) - 1 \right\} = 0 \text{ with } y(\varepsilon) = M(\varepsilon)b/d. \quad (13)$$

The extreme value of the quantity of PD, $\varepsilon_m$, under quasi-static loading follows from an approximate solution $y_0(\varepsilon)$ of the transcendental equation in the figure brackets (13), which depends from the parameter *ε*, and to be by the solution of the equation:

$$y_0(\varepsilon) = b\frac{Gb_\varepsilon^3}{2 \cdot k_B T} \Rightarrow \varepsilon_m(d,T) = f(y_0): \; \varepsilon_m(d,T) \equiv \sqrt[3]{y_0(\varepsilon_m)d/(M(0)b)} - 1, \quad (14)$$

Where a root $\varepsilon_m$ is chosen in the range $0 < \varepsilon_m < 1$. For $\sigma = \sigma(\varepsilon)$ of the form (1) the limit $\lim_{d \gg b} \varepsilon_m = 0,5$ holds.

---

[6] the property: $d_0(\varepsilon, T_1) > d_0(\varepsilon, T_2)$ for $T_1 < T_2$, is opposite to one marked by item 2) in the Introduction of Ref.[1] obtained within molecular dynamics simulations in [12] but for high .rate strain $\dot{\varepsilon}$.

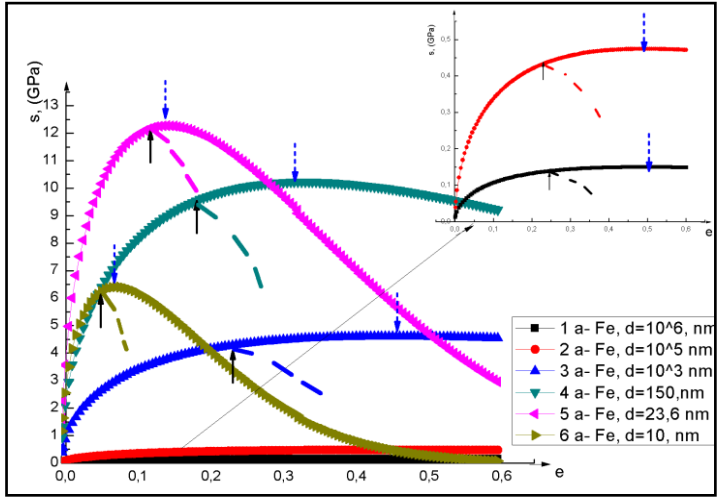

Fig. 3. Plots $\sigma = \sigma(\varepsilon,d,T)$ (1) for α-Fe at $T=300K$, $m_0 \cdot \alpha^2 = 3,66$, with stress-strain curves 1, 2, 3, 4, 5, 6 for $d=10^{-3}$; $10^{-4}$; $10^{-6}$m; $d=150$ nm: $d=d_0=23,6$ nm: $d=10$ nm for $T=300K$. By the black arrows below it is indicated the values, where the Backofen-Considére condition (15) is realized with $\varepsilon_{\text{fr.cond.}}$ (16) and expected dashed lines of curves before the fracture. The blue arrows from top indicate the maximums of $\sigma(\varepsilon)$ for the true strains calculated according (14). On the input the plots for the stress-strain curves for CG aggregates are shown.

When constructing the stress-strain curves we have used the Backofen-Considére condition of the fracture: $\sigma = d\sigma/d\varepsilon$ [3], which selects the regions of homogeneous and localized PD and permits one to determine the values of conditional strain $\varepsilon_{\text{fr.cond.}}$ and stress of fracture (ultimate stress) $\sigma_S$ from the equation:

$$\theta_0(\varepsilon) - \sigma_0(\varepsilon) = \alpha m \frac{Gb}{d\sqrt{\varepsilon}}\sqrt{\frac{3}{\pi\sqrt{2}}m_0 M(0)}\left(e^{M(\varepsilon)b/d}-1\right)^{-\frac{1}{2}} \times X(\varepsilon,d,T),$$
$$X(\varepsilon,d,T) = 2\varepsilon - 1 + \frac{3\varepsilon}{1+\varepsilon}M(\varepsilon)\frac{b}{d}e^{M(\varepsilon)b/d}\left(e^{M(\varepsilon)b/d}-1\right)^{-1}, \tag{15}$$

with account for (12), (13). In the approximation $\theta_0(\varepsilon) = \sigma_0(\varepsilon) = 0$ the solutions of (15) $\varepsilon_{\text{fr.cond.}}(d,T)$ are determined by the condition $X(\varepsilon,d,T) = 0$. For the values $d$ from the Table 4 we have for $\varepsilon_{\text{fr.cond.}}(d,300)$:

$$\varepsilon_{\text{fr.cond.}}(10; 23,6; 150; 10^3; 10^5; 10^6) = (0,04; 0,12; 0,19; 0,23; 0,245; 0,25). \tag{16}$$

The absolute maximal value $\sigma_{\max}(\varepsilon_m, d_m, 300) = 13,27$ GPa is numerically determined from the equations (1) and (14) with $d_m = 40,6$ nm and $\varepsilon_m = 0,2$. This value was not experimentally observed in the NC region, is close to the theoretical ultimate stress and is determined by the peculiarities of the model. Among the peculiarities one may select a single-mode property of the crystallites, not accounting for the second phase (using the terminology of Ref.[3]) with boundary grains – the regions between the pure crystallites, being filled by the crystallites of sizes $d_{GB} \ll d$ and by pores, considered, e.g. as accommodations of the nanopores.

### Two phase model for polycrystalline aggregate. Softening from the boundary grains

Overcoming of the above peculiarities is based, first, on the natural assumptions that an input from the second phase should be add additively to equilibrium $\sigma(\varepsilon)$ (1) with a proportionality coefficient $\kappa_1, 0 \leq \kappa_1 < 1$ (the value $\kappa_1 = 0$ means the absence of an explicit contribution from the boundary grains (GB) into hardening). Second, with increasing of an accumulated PD ε, a volume of intergrain regions is increased with changing of the second phase contents, and therefore the porous structure therein is increased, that leads on the final stage to appearance of cracks and destruction of the sample, as well as for sub-microcrystalline and NC materials this phase provides the slipping through pores of grains (or groups of grains) for sufficiently small PD, that it is by a previously unaccounted softening factor. We take into account the contribution from the GB in to the total stress of the sample by means of subtracting the stress in the porous part of the second phase of the aggregate with a coefficient of proportionality $\kappa_2, 0 \leq \kappa_2 \leq \kappa_1$. Discontinuity of the sample in areas of grain boundaries implies a necessity of an negative input from pores into total $\sigma(\varepsilon)$ in dependence from average size of such pores, being increased with accumulation of PD ε. Let's consider pores as the *formal crystallites* of average sizes $d_P$ of the same materials. Such model of two-phase system reminds one from the composite models variants [3]. We relate the pores in the material with the size of GB: considering that the more $d_P$ (and $d_{GB}$), the more the part of large-angle GB, and vice-versa the smaller $d_P$ (and $d_{GB}$), the part of small-angle GB is more.

One may roughly estimate the value of the parameter $\kappa_1$ as a part of a volume of the second phase of the sample around the crystallite with respect to the volume $V_C$ of the crystallite itself: $\kappa_1 = V_{GB}/V_C = (n_1 b)\pi d^2 / \frac{1}{6}\pi d^3 = n\frac{b}{d}$, with a some constant $n = 6n_1 \sim 10^0 - 10^2$, which takes into account the average distance between grains and highly depends on a preparation of the GB states. The constant $n$, as well appears, in general, by a function from average size, $d$, of the crystallites. The part of the volume of GB grows with decreasing of the crystallite diameter $d$, that means an increasing of a softening factor. The modified model, which takes into account for GB, including theirs pores structure, leads to the following form of dependence for the phase of the equilibrium integral flow stress:

$$\sigma_\Sigma(\varepsilon) = (1-\kappa_1)\sigma_C(\varepsilon,d) + (\kappa_1-\kappa_2)\sigma_{GB}(\varepsilon,d_{GB}) - \kappa_2\sigma_P(\varepsilon,d_P) \\ = (1-nb/d)\sigma_C(\varepsilon,d) + (n-m)(b/d)\sigma_{GB}(\varepsilon,d_{GB}) - m(b/d)\sigma_P(\varepsilon,d_P), \quad m \leq n, \qquad (17)$$

where $\sigma_C(\varepsilon,d) = \sigma(\varepsilon,d)$ is the stress for the first phase – being by the basic grains from the sample with diameter $d$; $\sigma_{GB}(\varepsilon,d_{GB})$ and $\sigma_P(\varepsilon,d_P)$ appear by the stresses for crystallites and pores from the GB region of the respective average sizes $d_{GB}$ and $d_P$. For $n=m$ all the area of the second phase is filled by the pores of different diameters. For $m=0$ the second phase represents a set of the crystallites of different sizes with $b < d_{GB} \ll d$. The last case appears by the model one. Let's choose uniform distribution both for the crystallites and for pores from the second phase with respect to its sizes in the range $[d_{\min} = n_{\min}b, d_{\max} = n_{\max}b]$, with integers $n_{\min} \leq n_{\max}$, to be experimentally determined as a result of the certification of GB states. If we assume, that $d_{GB} \leq 10^{-1}d$, we will have for $n_{\max} = d_0/b \sim 10^2$ with $d_0$ from (2) and choosing $n_{\min} = 5$, with absence of the grains from the second phase (i.e. with $n=m$) the FS of homogeneous two-phase polycrystalline material follows from the equations (1) and (17) in the form:

$$\sigma_\Sigma(\varepsilon) = \sigma_0(\varepsilon) + (1 - nb/d)\left\{\alpha m \frac{Gb}{d}\sqrt{\frac{6\sqrt{2}}{\pi}m_0 \varepsilon M(0)}\left(e^{M(\varepsilon)b/d} - 1\right)^{-\frac{1}{2}}\right\} - n(b/d)\sigma_P(\varepsilon), \qquad (18)$$

$$\sigma_P(\varepsilon) = \frac{1}{n_{\max} - 4}\sum_{i=5}^{n_{\max}}\sigma_P^{(i)}(\varepsilon), \quad \sigma_P^{(i)}(\varepsilon) = \alpha m \frac{G}{i}\sqrt{\frac{6\sqrt{2}}{\pi}m_0 \varepsilon M(0)}\left(e^{M(\varepsilon)/i} - 1\right)^{-\frac{1}{2}},$$

where a constant $\sigma_0(\varepsilon)$ is extracted from the first and second phases of the PC sample. For α-Fe, when $n=60$ (that corresponds to smaller-angle grain boundaries in the coarse-grain limit, but not in the NC region) the stress-strain curves for large $d$ (the plots 1 and 2 on the Fig. 3) is non-significantly decreased with respect to the stress (quantitatively corresponding to the results of the experiments on Armco-iron [11]), whereas in the region of PC materials, starting from UFG materials an negative input from the second phase with GB becomes by co-measurable one with input from the first phase (with basic crystallites), to be decreased both the values of the stress and narrowing the zone of plasticity – with the range of strain ε under PD up to fracture. In particular, for average size of pores $d_P = 150b$, taking from (18) the second term for $\sigma_P(\varepsilon)$ with $i = 150$, we get that for $d_m = 40,6$ nm the formal maximal value $\sigma_{\Sigma\max}(d_m,300) = (0,63*13,27 - 4,60) = 3,809$ GPa decreases in 3,5 times with respect to $\sigma_{\max}(d_m,300) = 13,27$ GPa, whereas the actual (for 2-phase model) maximum $\bar{\sigma}_{\Sigma\max}(d_m,300) = 3,812$ GPa is already reached for $\bar{\varepsilon}_m = 0,21 > \varepsilon_m$. The yield strength and the part of the second phase for the above-mentioned parameters of the materials are equal $\sigma_{\Sigma y|d_m} = \sigma_\Sigma(0,002) = 0,586$ GPa and 36%.

A detailed study of the flow stress (17), (18) with account for both crystallite part of the second phase and with arbitrary distributions with respect to sizes for the crystallites of the first and second phases together with analysis of an influence on FS of the GB states for BCC, FCC, HCP polycrystalline materials presents the subject of the further research.

## Conclusion.

In the paper the quasi-particle interpretation of the quantization of the energy of crystallite of PC aggregate under PD is given. Within this interpretation the quantum of energy of unit dislocation, condition-ally called as the *dislocon,* appears by the composite (short-lived) particle consisting from the acoustic phonons surrounding two atoms (at the moment of time of breaking the bond between the atoms and creating of a nanopore), which leave the nodes of the crystal lattice under PD. This idea allows one to come analytically to the description and evolution of the Chernov-Luders macroband of PD in connection with observed acoustic emission [12, 13]. The quasi-particle interpretation adds the significant argument in flavor of the concept of the origin of $1D$-defects (dislocations) in terms of $0D$-defects (nanopore, vacancy) [1] under PD process. It was shown within one-phase model of an single-mode PC material that with the grows of the temperature the maximum of the yield strength $\sigma_m$ increases in such a way that in the coarse-grained limit the yield strength, $\sigma_y$, decreases down to new critical size of the grain $d_1$: $d > d_1 \sim 3d_0$, whereas for the grains with $d < d_1$ the values of $\sigma_y$, including maximum $\sigma_m$, increase as it was theoretically demonstrated for pure Al on the Fig. 2. This new effect, which we called as the *temperature dimension effect*, should be modified when two-phase model will be applied (see [14] for small- and large GB angles and for special constant GB) and requires experimental verification (!). For the stress-strain curves for single-mode α-Fe (in assuming that there is no the second (GB) phase of the material) it was shown on the Fig. 3, that the stress maximum – ultimate strength $\sigma_S$ increases with decreasing of the linear grain size, shifting in to the region of smaller PD with value $\varepsilon_m$, from $\varepsilon_m = 0,5$ for CG material and reaches in the NC region of its absolute maximum with $(\varepsilon_m; d_m) = (0,2; 40,6$ nm$)$: $\sigma_{\max}(\varepsilon_m, d_m, 300) = 13,27$ GPa at $T=300$K. With decreasing of $d$ for the grain size the maximum of ultimate strength is significantly decreased together with plasticity region. The validity of Backofen-Considére criterion (15) is realized for $\varepsilon_{\text{fr.cond.}}(d)$ (16), making the maximums $\sigma_{\max}(d_m)$ by physically unreachable ones.

It was suggested the two-phase model of PC aggregate, which augments the model with only (solid) crystalline phase by means of introducing of the second phase, which describes the regions of grain boundaries. The latter includes both the crystalline (fragmentary) part with smaller sizes and the pores necessarily presented between the grains. The analytical presence and input of the second phase in to the phase of the equilibrium integral stress $\sigma_\Sigma(\varepsilon)$ (which has the dimension of the density of energy) was additively taken into account with positive input from the fragmentary and with negative one from the porous parts of the second phase. The relationships (17) and (18) with only porous structure of the second phase are by the basic results of the paper. The parameters in it correspond to the states and interrelations of the first and second phases of the PC aggregate. It was shown, on the example of α-Fe, that the stress-strain curves, and in particular, the value of maximum strongly depend upon the part of the GB in the aggregate and $\sigma_{\Sigma\max}(d_m, 300) = 3,809$ GPa decreases in 3,5 times with respect to $\sigma_{\max}(d_m, 300) = 13,27$ GPa for pure crystalline phase. The part of the second phase consists of 36% (in accordance with [5] and reference [14] cited in [1]), so that when transition to the single-mode sample with the size $d = d_0 = 23,6$ nm of the first phase crystallites is done, the parts of the phases in such sample are changed in 2 times being respectively reached to 30% and 70%, that means the impossibility of the material existence because of: $\sigma_\Sigma < 0$.

The development of the paper results for the materials with multimodal structure, on alloys, including additional dispersion hardening (e.g., for the steels see Ref. [15]) has an important application in the Strength and Plasticity of the Physics of polycrystalline metals. It is also necessary to find an equation leading to the energy spectrum of the crystallite.

The author is thankful to S.G. Psakhie, I.A. Dittenberg, V.V. Kibitkin, P.Yu. Moshin for the discussion of the paper results, to E.V. Shilko for support and discussions. The work has been done in the framework of the Program of fundamental research of State academies of Sciences for 2013-2020.